\documentclass{pasj00}

\newcommand{\av}{\ensuremath{A_{V}}}

\newcommand{\kt}{\ensuremath{k_{\rm{B}}T}}
\newcommand{\lx}{\ensuremath{L_{\rm{X}}}}

\newcommand{\fx}{\ensuremath{F_{\rm{X}}}}
\newcommand{\nh}{\ensuremath{N_{\rm H}}}

\SetRunningHead{Tsujimoto, Hyodo, \& Koyama}{Suzaku Spectroscopy of Arches Cluster}
\begin{document}
\title{Suzaku Spectroscopy Study of Hard X-Ray Emission\\ in the Arches Cluster}
\author{Masahiro \textsc{Tsujimoto}}
\affil{Department of Physics, Rikkyo University, 3-34-1, Nishi-Ikebukuro, Toshima, Tokyo 171-8501.}
\email{tsujimot@tvs.rikkyo.ne.jp}
\author{Yoshiaki \textsc{Hyodo} and Katsuji \textsc{Koyama}}
\affil{Department of Physics, Graduate School of Science, Kyoto University,\\ Kita-shirakawa Oiwake-cho, Sakyo, Kyoto 606-8502.}
\email{hyodo@cr.scphys.kyoto-u.ac.jp, koyama@cr.scphys.kyoto-u.ac.jp}
\maketitle

\begin{abstract}
 We present the results of a Suzaku study of the Arches cluster in the Galactic center
 region. A high signal-to-noise ratio spectrum in the 3--12~keV band was obtained with
 the XIS (X-ray Imaging Spectrometer) onboard the Suzaku Observatory. We found that the
 spectrum consists of a thermal plasma, a hard power-law tail, and two Gaussian line
 components. The plasma component with a temperature of $\sim$2.2~keV is established
 from the presence of Ca\emissiontype{XIX} and Fe\emissiontype{XXV}
 K$\alpha$ lines as well as the absence of Fe\emissiontype{XXVI} K$\alpha$
 line. The two Gaussian lines represent the K$\alpha$ and K$\beta$ lines from iron at
 lower ionization stages at $\sim$6.4 and $\sim$7.1~keV. Both the line centers and the
 intensity ratio of these two lines are consistent with the neutral iron. The hard
 power-law tail with a photon index of $\sim$0.7 was found to have no pronounced iron K
 edge feature. In comparison with the published Chandra spectra constructed separately
 for point-like and diffuse emission, we conclude that the thermal component is from the
 ensemble of point-like sources plus thermal diffuse emission concentrated at the
 cluster center, while the Gaussian and the hard tail components are from the
 non-thermal diffuse emission extended in a larger scale. In the band-limited images of
 the XIS field, the distribution of the 7.5--10.0~keV emission resembles that of the
 6.4~keV emission, including the local excess at the Arches cluster. This strongly
 suggests that the power-law emission is related to the 6.4 and 7.1~keV lines in the
 underlying physics. We discuss two ideas to explain both the hard continuum and the
 lines: (1) X-ray photoionization that produces fluorescence lines and the Thomson
 scattering continuum and (2) non-thermal electron impact ionization of iron atoms and
 bremsstrahlung continuum. But whichever scenario is adopted, the photon or particle
 flux from the Arches cluster is too low to account for the observed line and continuum
 intensity.
\end{abstract}

\section{Introduction}
The central $\sim$200~pc of the center of our Galaxy is particularly rich in curious
spatial and spectral features.  One of the features distinctively seen in the Galactic
center region is the diffuse iron K$\alpha$ line emission at $\sim$6.4~keV
\citep{koyama96}. The line originates from neutral or low-ionized iron (hereafter quoted
as Fe\emissiontype{I}), but its ionizing mechanism is not yet understood. Two major
ideas have been developed. One is fluorescence from photo-ionized iron
\citep{sunyaev93,koyama96,revnivtsev04}. The X-ray spectrum from the Sagittarius B2
cloud, which is characterized by the strong $\sim$6.4~keV emission line and the
$\sim$7.1~keV iron K edge, is best accountable by this model
\citep{koyama96,murakami00}. Another idea is that iron is ionized by accelerated
electrons \citep{yusef-zadeh02b,predehl03,wang06}.

The distribution of the diffuse 6.4~keV emission is highly asymmetric around the
Galactic center with a bias in the north-east direction \citep{koyama96,koyama06b} and
appears filamentary \citep{park04}. Of particular interest is the 6.4~keV emission
associated with the Arches cluster, which is one of the richest and the most densely
packed massive star clusters in our Galaxy.

The cluster has a total mass of $\sim$10$^{4}$~$M_{\odot}$, a compact size with a
diameter of $\sim$15\arcsec\ ($\sim$0.6~pc at an assumed distance of 8.5~kpc), a stellar
mass density of $\sim$3$\times$10$^{5}$~$M_{\odot}$~pc$^{-3}$, and an estimated age of
2--4.5~Myr
\citep{nagata95,cotera96,serabyn98,figer99,blum01,yang02,stolte02,figer02,figer05}. It
contains $\sim$5\% of all the known Wolf-Rayet stars in our Galaxy \citep{figer05},
which are characterized by a large mass loss rate reaching
$\sim$10$^{-4}$~$M_{\odot}$~yr$^{-1}$ \citep{lang01,lang05}. The level of high mass
star-forming activity of the Arches cluster, which can be measured by the number of O
stars, is comparable only to NGC\,3603 \citep{moffat83}, W49A \citep{conti02,alves03},
and Westerlund 1 \citep{clark05} in our Galaxy and R136 at the center of 30 Doradus
\citep{campbell92} in the Large Magellanic Cloud.

\citet{yusef-zadeh02} first reported the detection of two point sources in addition to
diffuse emission from this region using the Chandra X-ray Observatory
\citep{weisskopf02}. The extended X-ray emission is elongated beyond the cluster
boundary with a size of $\sim$60\arcsec$\times$90\arcsec\ and its spectrum can be fit
with a thermal model with a temperature of $\sim$5.7~keV in addition to a $\sim$6.4~keV
line \citep{yusef-zadeh02}. Based on these claims, the hard diffuse emission is
considered an example of the cluster wind plasma, which is a collection of interacting
stellar winds from massive stars in a cluster
\citep{canto00,raga01,silich04,rockefeller05}. However, the poor spectrum of the
extended emission does not allow a conclusion to be drawn about a thermal
origin. \citet{law04} cautioned that the continuum emission can also be fit by a
power-law, and can thus be explained by non-thermal emission. Recent results based on a
longer exposure Chandra observation by \citet{wang06} indicated that the cluster wind is
only seen at the cluster center, and the majority of the diffuse emission is composed of
a strong 6.4~keV line and a power-law continuum emission of non-thermal origin.

In order to understand the nature of the X-ray emission in the Arches cluster and the
diffuse 6.4~keV emission ubiquitous in, but exclusive to, the Galactic center region, 
hard-band X-ray spectroscopy with high signal-to-noise ratio (S/N) is
obviously a key. The bandpass includes both thermal (Fe\emissiontype{XXV} K$\alpha$ and
Fe\emissiontype{XXVI} K$\alpha$ at $\sim$6.7 and $\sim$7.0~keV) and fluorescent
(Fe\emissiontype{I} K$\alpha$ and K$\beta$ at $\sim$6.40 and $\sim$7.05~keV) features as
well as the iron K absorption edge feature at $\sim$7.11~keV. The underlying power-law
continuum, if it ever exists, is visible at energies $\gtrsim$~8~keV, where the thermal
contribution plays a minor role. Moreover, the hard X-rays penetrate through the large
attenuation of \av$\sim$30~mag common toward the Galactic center region
\citep{morris96}.

Here, we present the results of a spectroscopic study of the hard emission in the Arches
cluster using the Suzaku Observatory. The XIS (X-ray Imaging Spectrometer;
\cite{koyama06a}) onboard Suzaku \citep{mitsuda06} provides high S/N spectra in the
hard-band aided by the excellent spectroscopic performance of the CCDs and the large
effective area of the X-ray optics. To compliment the moderate spatial resolution of
Suzaku, we refer to the recent Chandra results by \citet{wang06} and results obtained by
our own reduction of the same dataset.

\section{Observations and Data Reduction}\label{s3}
Two observations of the Galactic center region were conducted using Suzaku during the
performance verification phase on 2006 September 23 and 30. Suzaku observations provide
simultaneous XIS and HXD (Hard X-ray Detector; \cite{takahashi06,kokubun06}) data. We
concentrate on the XIS data with a combined integration time of $\sim$95~ks. The XIS
field is centered at (R.\,A., Decl.)$=$ (\timeform{17h46m03s}, \timeform{-28D55'32''})
in the equinox J2000.0. The off-axis angle of the Arches cluster is $\sim$6\farcm6.

The XIS is equipped with four X-ray CCDs. Three of them (XIS0, 2, and 3) are
front-illuminated (FI) CCDs and the remaining one (XIS1) is a back-illuminated (BI) CCD
that is superior in the soft band responses. They are mounted at the focus of four
independent X-ray telescopes (XRT; \cite{serlemitsos06}). The detectors are sensitive in
the energy range of 0.2--12.0~keV with an initial energy resolution of $\sim$130~eV in
the full width half maximum\footnote{The spectral resolution keeps degrading in the
orbit. We included a degradation of $\sim$30~eV (1 $\sigma$) for the presented analysis,
which was measured from the same dataset \citep{koyama06b}.} at 5.9~keV ($R\sim$~45) and a
total effective area of $\sim$590~cm$^{2}$ at 8~keV. An XIS field of view covers a
$\sim$18\arcmin $\times$18\arcmin\ region with a half power diameter (HPD) of
$\sim$1\farcm9. The HPD is almost independent of the off-axis angle within $\sim$10\%
\citep{shibata01}.

Data were taken by the XIS normal mode and screened to remove events during the South
Atlantic Anomaly passages and earth elevation angles below 5 degrees. There are standard
radioactive sources of \atom{Fe}{}{55} installed in two corners of each of the four
CCDs. These calibration sources were used to determine the absolute gain. In addition,
we can utilize the observed intense emission lines from the hot gas filling the Galactic
center region \citep{koyama06b}. Assuming that the center energy of Fe and S K$\alpha$
lines in the diffuse spectrum is spatially uniform, we corrected for the charge transfer
inefficiency and fine-tuned the energy gain across the chips. These calibration
processes were performed using the same dataset with the presented
analysis. Resultantly, the systematic energy uncertainty of XIS is as low as $\sim$
$+$3/$-$6~eV at 6~keV. The details of the calibration of the Galactic center datasets
are described in \citet{koyama06b}.

\section{Analysis}\label{s4}
\subsection{Image}
\begin{figure}
 \begin{center}
  \FigureFile(75,75){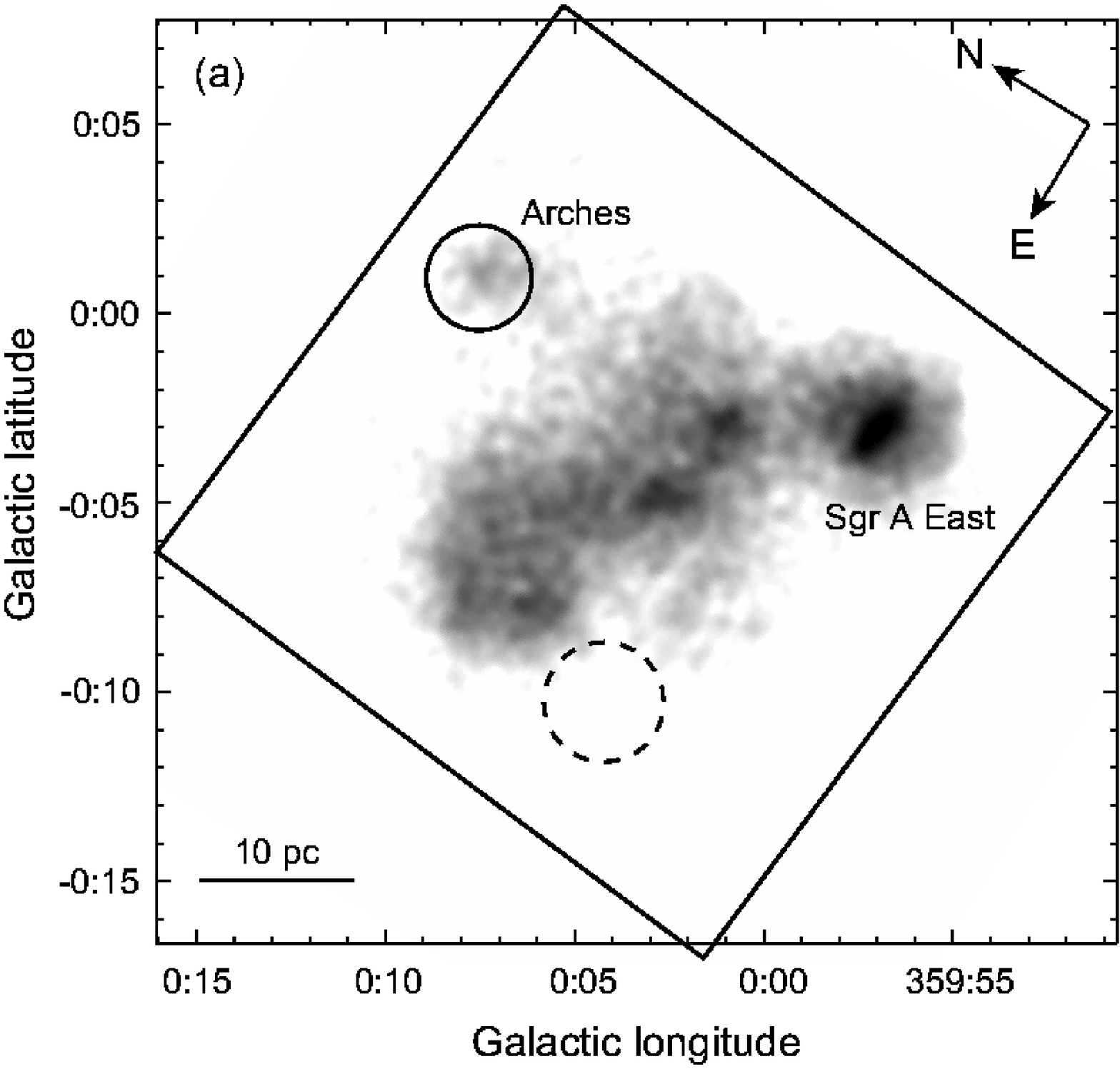}
  \FigureFile(75,75){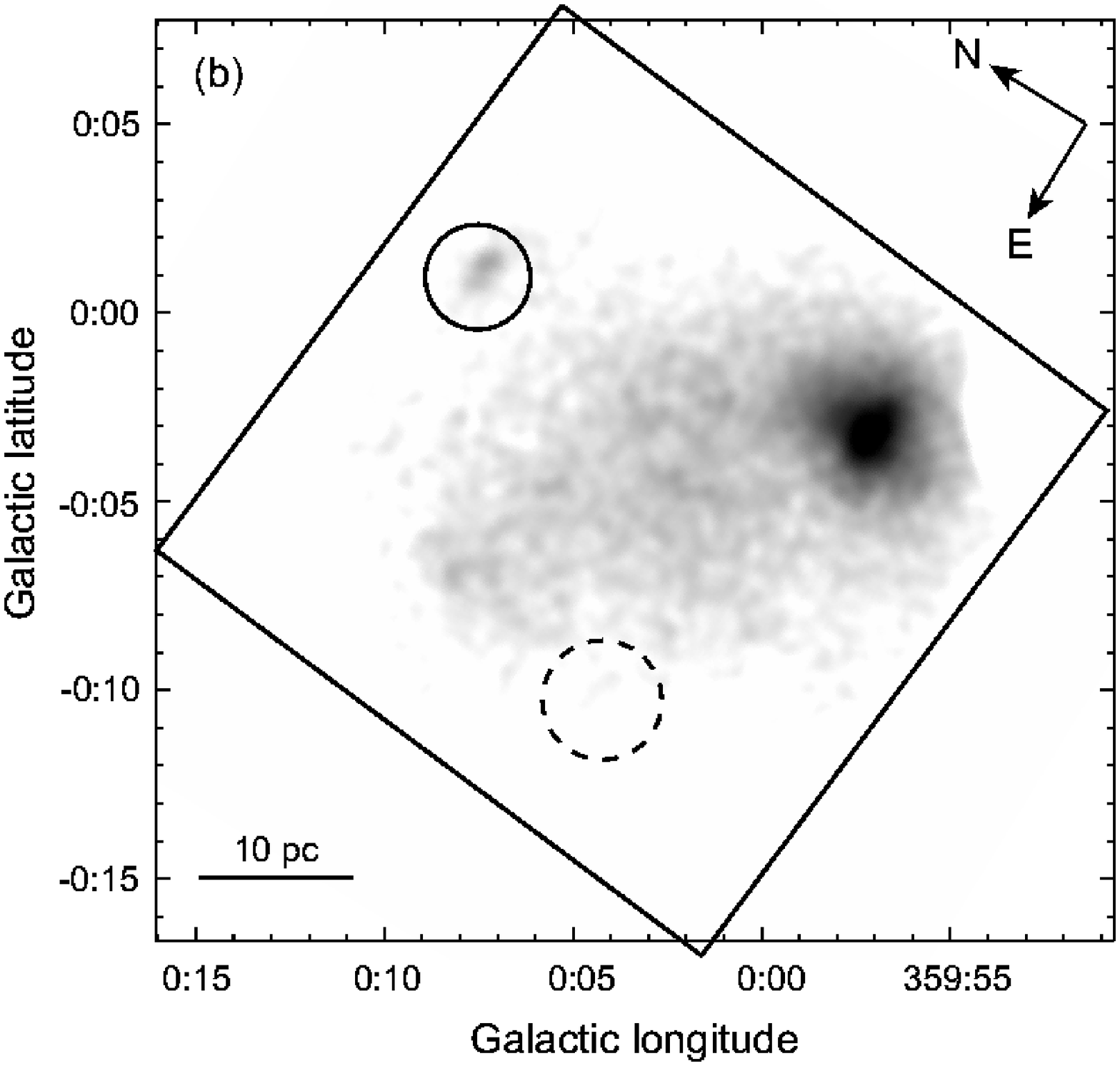}
  \FigureFile(75,75){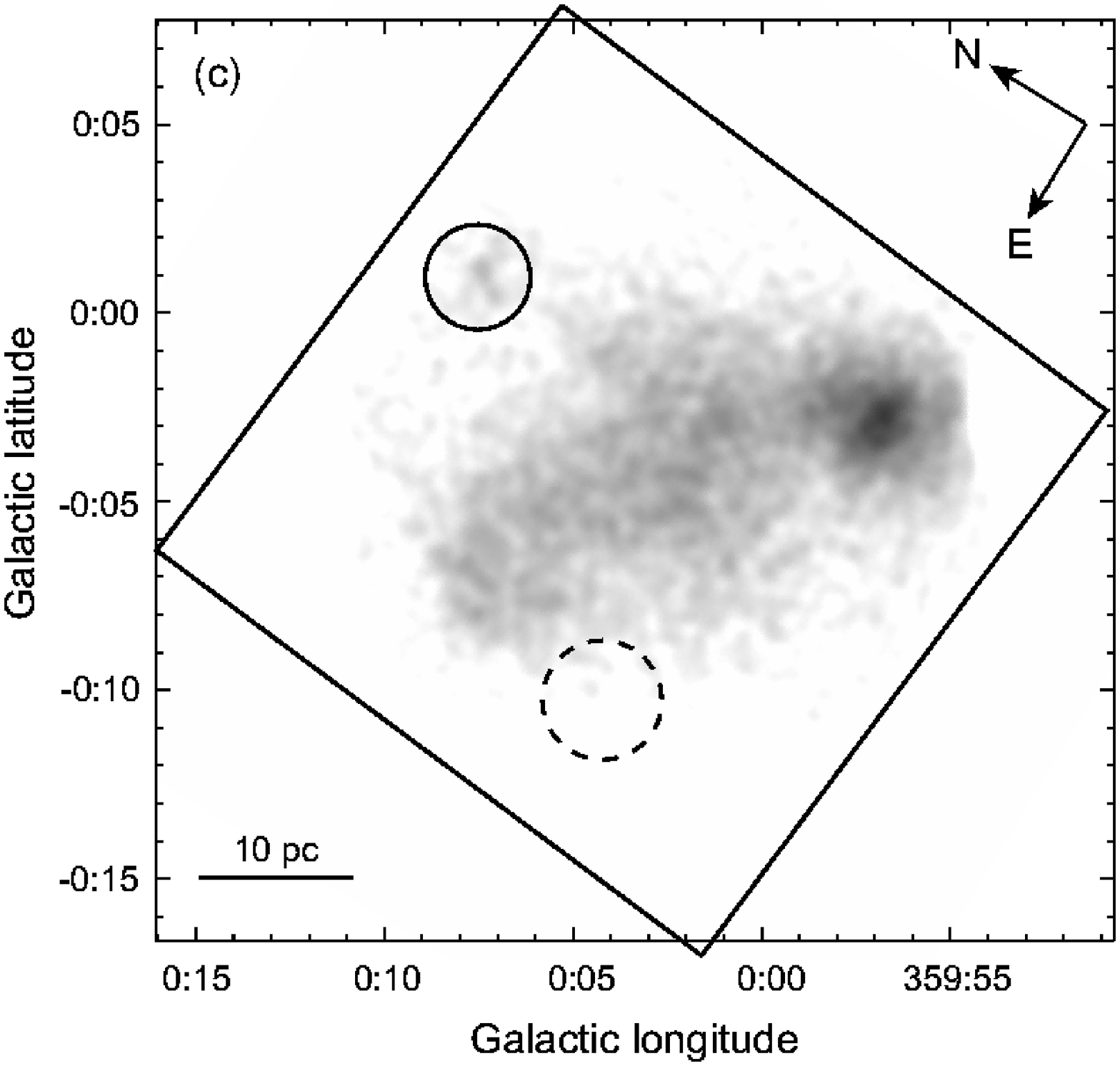}
 \end{center}
 \caption{Band-limited smoothed X-ray images by XIS: (a) 6.25--6.55, (b) 6.55--6.85, and
 (c) 7.5--10.0~keV ranges. The data from the four CCDs are merged. Solid and dotted
 circles show the source and background accumulation regions, respectively. The four
 corners of the images are masked out to remove signals from calibration
 sources.}\label{fg:f1}
\end{figure}

Figure~\ref{fg:f1} shows the band-limited XIS images in the (a) 6.25--6.55, (b)
6.55--6.85, and (c) 7.5--10.0~keV bands. The former two bands contain K$\alpha$ emission
lines from iron at low and high ionization stages, respectively. The 7.5--10.0~keV image
complements the one below 8.0~keV by Chandra; the effective area at 8~keV of Suzaku XIS
is larger than that of Chandra ACIS (Advanced CCD Imaging Spectrometer;
\cite{garmire03}) by more than ten-fold. A significant excess is seen at the position of
the Arches cluster in all bands, which is elevated from the diffuse emission at the
north east of the Galactic center. Due to the limited spatial capability of Suzaku XIS,
it is difficult to discriminate point sources from the underlying diffuse emission in
the Arches cluster.

The 7.5--10.0~keV image (Fig.~\ref{fg:f1}c) morphologically resembles the 6.4~keV image
(Fig.~\ref{fg:f1}a) more than the 6.7~keV image (Fig.~\ref{fg:f1}b). In particular, both
the 7.5--10.0~keV and the 6.4~keV images have a local excess centered at the radio arc
bubble located at about ($l$, $b$)$=$(0.13\arcdeg, --0.11\arcdeg). The correlation
between the 6.4~keV emission and the radio arc bubble was found by
\citet{rodriguez01}. Also, the 6.7~keV image has the dominating peak at the Saggitarius
A East, while the 6.4~keV and 7.5--10.0~keV images have extended emission across the
field with a similar surface brightness.

\subsection{Spectrum}
\begin{figure}
 \begin{center}
  \FigureFile(85mm,50mm){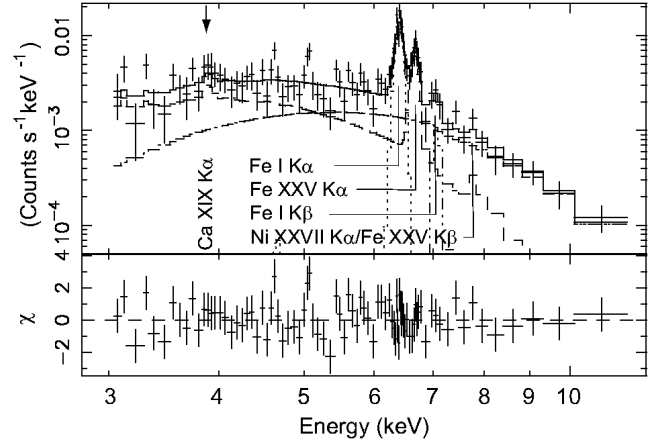}
 \end{center}
 \caption{XIS spectrum of the Arches cluster composed of point-like and diffuse
 emission. Grouped and background-subtracted data with uncertainties are plotted in the
 upper panel. The best-fit model convolved with the mirror and detector responses are
 shown by the broken (each component) and solid (total) lines. The lower panel shows the
 residuals of the fit.}\label{fg:f2}
\end{figure}

Figure~\ref{fg:f2} shows the background-subtracted XIS spectrum of the Arches cluster
constructed from the three FI chip data. The source spectrum was extracted from a circle
around the local intensity peak of a smoothed XIS image (solid circle in
Fig.~\ref{fg:f1}), while the background was accumulated from a position devoid of
intense diffuse emission and at a similar off-axis angle (dashed circle in
Fig.~\ref{fg:f1}). We tried several different positions for the background subtraction
and found that the spectrum above $\sim$3~keV is insensitive to the background
positions. The spectrum below $\sim$3~keV is contaminated by the inhomogeneous diffuse
emission typical to the Galactic center region, hence is not used hereafter. A radius of
1\farcm6, which is the $\sim$80\% encircled energy radius of a point source, is used for
the background extraction circle. A smaller radius of 1\farcm4 is used for the source
circle to maximize the S/N.

The spectrum is characterized by a complex of iron lines in the 6.0--8.0~keV band over a
hard continuum. The three most conspicuous lines are attributable to the K$\alpha$ line
from highly ionized iron (Fe\emissiontype{XXV}) at $\sim$6.7~keV, and K$\alpha$ and
K$\beta$ lines from neutral or low-ionized iron (Fe\emissiontype{I}) at $\sim$6.4 and
$\sim$7.1~keV, respectively. The existence of Fe\emissiontype{XXV} K$\alpha$ in addition
to Ca\emissiontype{XIX} K$\alpha$ ($\sim$3.9~keV) and Ni\emissiontype{XXVII} K$\alpha$
$+$ Fe\emissiontype{XXV} K$\beta$ (7.8--7.9~keV) lines indicates a thermal plasma with a
temperature of a few keV, while Fe\emissiontype{I} K lines and a hard continuum up to
$\sim$12~keV suggest additional components.

We first fitted the spectrum with photo-electrically absorbed thin-thermal plasma (APEC;
\cite{smith01}) model to determine the plasma temperature (\kt) and the amount of
interstellar extinction (\nh). The abundance relative to solar was derived separately
for elements with prominent emission lines. Two Gaussian components were added to the
model to account for the lines at $\sim$6.4 and $\sim$7.1~keV. Their line center
($E_{1}$ and $E_{2}$) and photon flux ($N_{1}$ and $N_{2}$) were also determined by the
fit. The widths of the lines were consistent with zero.

The resultant best-fit model was marginally acceptable with the null hypothesis
probability of the $\chi^{2}$ value ($P_{\chi^{2}}$) of 0.01. A significant residual above
$\sim$7~keV obviously requires an additional component. We added a power-law component
to reach an acceptable fit. An additional thermal component also yielded an acceptable
fit, but the best-fit temperature of $>$20~keV is unphysical. We therefore continued
with the power-law component.

The final fit result ($P_{\chi^{2}}=$0.28) is summarized in Table~\ref{tb:t2}. Here, we
fixed the abundance to be solar for all the elements upon the confirmation that the
best-fit value is consistent with solar. Also, $E_{2}$ and $N_{2}$ are tied to $E_{1}$
and $N_{1}$ respectively, as the best-fit values are consistent with the ones expected
for the neutral iron K$\alpha$ and K$\beta$ lines. By separating the Fe\emissiontype{I}
K$\beta$ line at $\sim$7.05~keV, the Fe\emissiontype{XXVI} K$\alpha$ line at
$\sim$6.97~keV and the Fe K edge at $\sim$7.11~keV are found to be absent. The absence
of the Fe\emissiontype{XXVI} line as well as the existence of the Fe\emissiontype{XXV}
line contributes for a tight constraint of the plasma temperature. The \nh\ is common
for all the components. The X-ray flux and luminosity (\fx\ and \lx) in the
3.0--10.0~keV band were calculated separately for the thermal and the power-law
components. The \lx\ value is corrected for the absorption with the distance to the
source assumed to be 8.5~kpc.

 \begin{table*}
 \caption{XIS Spectroscopy.}\label{tb:t2}
 \begin{center}
  \begin{tabular}{llll}
   \hline
   \hline
   Components & Par. & Units & Values\footnotemark[$*$]\\
   \hline
   Extinction & \nh & cm$^{-2}$ & 14 (9.1--19) $\times$10$^{22}$~\\
   Thermal   & \kt & keV & 2.2 (1.6--3.8) \\
             & \fx\footnotemark[$\dagger$] & erg~s$^{-1}$~cm$^{-2}$ & 5.2 (2.0--13) $\times$ 10$^{-13}$ \\
             & \lx\footnotemark[$\dagger$] & erg~s$^{-1}$ & 8.8 (3.4--21) $\times$ 10$^{33}$ \\
   Power-law & $\Gamma$ & & 0.72 (0.0--1.4) \\
             & \fx\footnotemark[$\dagger$] & erg~s$^{-1}$~cm$^{-2}$ & 7.3 (1.6--28) $\times$ 10$^{-13}$ \\
             & \lx\footnotemark[$\dagger$] & erg~s$^{-1}$ & 9.3 (2.0--36) $\times$ 10$^{33}$ \\
   Gaussian  & $E_{1}$ & keV & 6.41 (6.40--6.42) \\
             & $N_{1}$ & s$^{-1}$~cm$^{-2}$ & 2.1 (1.8--2.5) $\times$10$^{-5}$ \\
             & $E_{2}$ & keV & 7.07 $=$ 1.103 $\times E_{1}$ \\
             & $N_{2}$ & s$^{-1}$~cm$^{-2}$ & 2.4 $\times$ 10$^{-6}$ $=$ 0.113$\times N_{1}$\\
   \hline
   \multicolumn{4}{@{}l@{}}{\hbox to 0pt{\parbox{160mm}{
   \footnotesize
   \footnotemark[$*$] Uncertainties in the paralentheses indicate the 90\% confidence range.
   \par\noindent
   \footnotemark[$\dagger$] Values in the 3.0--10.0~keV band.
   }\hss}}
  \end{tabular}
 \end{center}
\end{table*}

\section{Discussion}\label{s5}
\subsection{Origin of Three Spectral Components}
We identified three spectral components (thermal, power-law, and two Gaussian lines) in
the XIS spectrum. \citet{wang06} conducted spatially-resolved Chandra spectroscopy of
the Arches cluster in the 2--8~keV band and revealed a more detailed spatial structure
upon the results by \citet{yusef-zadeh02,law04}. The Arches X-rays consist of three
spatial components: (1) point-like sources with a thermal spectrum, (2) diffuse emission
with a thermal spectrum in a $\sim$30\arcsec\ scale, and (3) diffuse emission with a
$\sim$6.4~keV line over a power-law continuum in a more extended scale
($\sim$60\arcsec$\times$90\arcsec\ by \cite{yusef-zadeh02}). The former two thermal
components are concentrated at the cluster center, in which three brightest point
sources dominates the total emission. The last non-thermal component extends toward the
southeast direction of the cluster. We consider that the thermal component in the XIS
spectrum is from the ensemble of point sources plus the thermal diffuse emission at the
cluster center, while the $\sim$6.4~keV and the accompanying $\sim$7.1~keV lines as well
as the power-law continuum are from the larger diffuse emission.

In order to compare directly the best-fit spectral model parameters between the ACIS and
XIS spectra, we reduced the Chandra ACIS data and obtained the best-fit parameters for
the composite of the three brightest point sources and the extended emission. The
parameters from XIS and ACIS spectral analysis agree with each other with overlapping
90\% confidence ranges. For the thermal component, the XIS spectrum has a temperature of
2.2 (1.6--3.8)~keV and the APEC normalization of 3.2 (1.2--7.8) $\times$
10$^{-3}$~s$^{-1}$~cm$^{-2}$~keV$^{-1}$, while the ACIS spectrum has 2.0 (1.9--4.4)~keV
and 2.5 (1.8--3.0) $\times$ 10$^{-3}$~s$^{-1}$~cm$^{-2}$~keV$^{-1}$. For the power-law
component, the photon index in the XIS spectrum is 0.72 (0.0--1.4), while that in the
ACIS diffuse spectrum is 1.2 (0.80--1.8). For the Gaussian lines, the photon flux of the
$\sim$6.4~keV line is 2.1 (1.8--2.5) $\times$10$^{-5}$~s$^{-1}$~cm$^{-2}$ in XIS and 2.3
(2.0--2.9) $\times$10$^{-5}$~s$^{-1}$~cm$^{-2}$ in ACIS. The equivalent width of the
line against the power-law component is $\sim$1.42~keV in XIS and $\sim$1.25~keV in
ACIS, indicating that the power-law flux as well as the line flux is consistent between
the two.

\subsection{Diffuse Medium}
The high S/N spectra obtained with the XIS give a stringent constraint on the center
energy of the $\sim$6.4~keV line, which is an increasing function of the ionization
stage of iron. The fluorescent line energy is also affected by whether and how the atom
is bound in molecules, but the resultant shifts are negligible of $\sim$1~eV
\citep{paerels98}. The best-fit value is consistent with the K$\alpha$ line from neutral
iron (6.40~keV). This is in agreement with the general understanding that most of iron
in the ISM is in the form of dust \citep{sofia94}. However, slightly low ionization
stages are also allowed up to about fourteenth (Fe\emissiontype{XV}; \cite{house69}),
including the systematic uncertainty of the XIS energy gain of $\sim$ $+$3/$-$6~eV.

\subsection{Cause of Line and Power-law Emission}
From the XIS spectrum (Fig.~\ref{fg:f2}), we found that the power-law component extends
up to $\sim$12~keV and dominates the spectrum above $\sim$8~keV. From the band-limited
XIS images (Fig.~\ref{fg:f1}), we noticed a similarity in the spatial distribution
between the 7.5--10.0~keV and the 6.4~keV emission. This indicates that the power-law
emission is related to the 6.4~keV line in the underlying physical process. Therefore,
the power-law component and the Fe\emissiontype{I} lines need to be explained
simultaneously. The equivalent width of the lines and the normalization of the power-law
comprise two major observational tests to discriminate ideas on the origin of the
emission. We examine two mechanisms (photoelectric ionization and electron impact
ionization) and derive the conditions for the primary ionizing beam. In the former, the
Thomson scattering continuum and the fluorescence are respectively responsible for the
power-law and the line emission. In the latter, non-thermal electron bremsstrahlung and
the K shell vacancy filling after the electron ionization are considered.

\subsubsection{Photoelectric Ionization}
We first note that the lack of a prominent absorption edge feature at $\sim$7.11~keV
alone does not constitute evidence against photoionization \citep{revnivtsev04}. This is
expected when the reflecting matter is optically-thin to the primary radiation. In fact,
the optically-thin approximation is justified by the Chandra result on the Arches
cluster, wherein the power-law emission suffers an extinction accountable only by the
ISM extinction to the Galactic center region ($\sim$6$\times$10$^{22}$~cm$^{-2}$;
\cite{wang06}).

We assume that the observed continuum emission does not include the direct X-rays. If it
does, the equivalent width of the iron K$\alpha$ fluorescent line is expected to be
$\lesssim$180~eV \citep{reynolds03,tsujimoto05} for the solar abundance, which
contradicts the observed value. The equivalent width of the fluorescent line against the
Thomson scattering continuum is expressed as a function of the ratio between the
photoelectric and Thomson scattering cross sections, the ratio of the target (electron
and iron atom) densities, and the geometry; the Thomson scattering is angle-dependent
while the fluorescence is spherically symmetric \citep{liedahl98}. Assuming that the
incident X-ray spectrum is $I(E)$, the electron and iron densities of the reflecting
medium are uniform ($n_{\mathrm{e}}$ and $n_{\mathrm{Fe}}$, respectively), the
photoelectric absorption cross section by iron is
$\sigma_{\mathrm{P}}^{\mathrm{Fe}}(E)$, the differential Thomson scattering cross
section is $(d\sigma_{\mathrm{T}}/d\Omega)(\theta)$ where $\theta$ is the angle between
the incident and scattered X-rays, the K$\alpha$ fluorescence yield is
$Y_{\mathrm{K}\alpha}$, and the K edge energy is $\chi$, the equivalent width of the
K$\alpha$ line is given by
\begin{eqnarray}
 \mathrm{EW}_{\mathrm{K}\alpha} &=& Y_{\mathrm{K}\alpha} \left(\frac{n_{\mathrm{Fe}}}{n_{\mathrm{e}}}\right) \left(4\pi \frac{d\sigma_{\mathrm{T}}}{d\Omega}(\theta)\right)^{-1}\nonumber\\
  &\times& \left(\frac{\int_{\chi}^{\infty}dEI(E)\sigma_{\mathrm{P}}^{\mathrm{Fe}}(E)}{I(E=6.4~\mathrm{keV})}\right).
\end{eqnarray}
By substituting $Y_{\mathrm{K}\alpha} \sim$0.34 \citep{kortright01},
$n_{\mathrm{Fe}}/n_{\mathrm{e}} \sim n_{\mathrm{Fe}}/n_{\mathrm{H}} \sim 3\times10^{-5}$
\citep{daeppen00} where $n_{\mathrm{H}}$ is the hydrogen density,
$d\sigma_{\mathrm{T}}/d\Omega (\theta) \sim 4.0 \times 10^{-26}
(1+\cos^{2}{\theta})$~cm$^{2}$, $I(E) \propto E^{-\Gamma}$, and
$\sigma_{\mathrm{P}}^{\mathrm{Fe}}(E) = 2 \times 10^{-20}
(E/7.1~\mathrm{keV})^{-3}$~cm$^{2}$ \citep{gullikson01}, we obtain
\begin{eqnarray}
\mathrm{EW}_{\mathrm{K}\alpha} &\sim& 3 \left(\frac{1}{\Gamma+2}\right)\left(\frac{6.4}{7.1}\right)^{\Gamma}\left(\frac{1}{1+\cos^{2}{\theta}}\right)~\mathrm{keV}.
\end{eqnarray}
We can determine both $\mathrm{EW}_{\mathrm{K}\alpha}$ and $\Gamma$ observationally,
resulting in $\theta \approx $~50~degrees using the best-fit values. In practice,
though, higher accuracy in the parameter determination is required for a meaningful
geometrical constraint. The observed $\mathrm{EW}_{\mathrm{K}\alpha}$ is easily
explained by the photoionization interpretation. For the amplitude of continuum
emission, the incident flux of $I(E) \sim 3 \times 10^{7}
(n_{\mathrm{e}}/10^{2}~\mathrm{cm}^{-3})^{-1}$~s$^{-1}$~cm$^{-2}$~keV$^{-1}$ is required
at 10~keV. Here, the diffuse medium is assumed to have a spherical shape of a 3~pc
diameter. The estimate of the electron density is from \citet{rodriguez01}.

To summarize, the photoionization interpretation requires the following conditions: (1)
the primary source is an external source, (2) the incident spectrum has a power of
$\Gamma \sim$~1 and (3) a photon flux of $\sim 10^{7}$~s$^{-1}$~cm$^{-2}$~keV$^{-1}$ at
10~keV, (4) the direct emission is optically thin to the reflecting medium (i.e.,
$n_{\mathrm{H}} < 2 \times 10^{5}$~cm$^{-2}$), and (5) the reflecting geometry should
satisfy the constraint on $\theta$.

No source is found for the primary source in the vicinity at present. The point sources
in the Arches cluster can be excluded. This is simply because the continuum emission by
the Thomson scattering cannot exceed the illuminating thermal emission above the iron K
edge energy, which is contrary to the XIS spectrum showing stronger non-thermal emission
than thermal emission at the band (Fig.~\ref{fg:f2}). \citet{wang06} also discussed that
Arches point sources are too dim for the fluorescent lines.

\subsubsection{Electron Ionization}
Another favored interpretation for the cause of the $\sim$6.4~keV line is the vacancy
filling after the iron K shell ionization by low energy electrons with an energy of
10--100~keV. This was proposed to explain the Galactic ridge diffuse X-ray emission
\citep{valinia00}, in which they claimed that the low energy electrons with a density of
$\sim$0.2~eV~cm$^{-3}$ and the power-law spectrum of an index of $\sim$0.3 contribute
significantly to both the observed line and continuum emission. \citet{yusef-zadeh02b}
and \citet{wang06} employed this model to account for the $\sim$6.4~keV line observed in
G0.13--0.13 and the Arches cluster in the Galactic center region, respectively.

The calculation of expected line and continuum intensity of this process requires
numerical treatments because of the complex dependence of the cross sections on the
electron and X-ray energy. A detailed computation is given in \citet{tatischeff02}. Two
important things have to be considered. First, the electron beam is stopped at the
surface of the hydrogen column. The stopping range for 10 and 100~keV electrons is
$\sim$7$\times$10$^{19} m_{\mathrm{H}}$ and $\sim$4$\times$10$^{21}
m_{\mathrm{H}}$~g~cm$^{-2}$ where $m_{\mathrm{H}}$ is the hydrogen mass
\citep{tatischeff02}. Second, the energy conversion rate from the electron beam to the
X-ray bremsstrahlung is quite small with an order of $\sim$10$^{-5}$.

With these kept in mind, we can compare the expected and observed values. The equivalent
width of the iron K$\alpha$ line against the Bremsstrahlung continuum is expected to be
$\sim$290~eV \citep{tatischeff02}, which is smaller than the observed value. An iron
abundance of 4--5 times larger than the solar value needs to be introduced to reconcile
the discrepancy, though no such indication is present for the Arches cluster. From the
spectral fit of the XIS spectrum, we found that a solar abundance for iron accounts for
the observed Fe\emissiontype{XXV} K$\alpha$ line intensity over the
continuum. \citet{wang06} suggested that the iron abundance is $\sim$2 from the
X-ray spectral fits of point sources in the cluster. \citet{najarro04} conducted
near-infrared spectroscopy of the photospheric emission of Wolf-Rayet stars in the
Arches cluster and found that the abundance is consistent with solar.

For the level of continuum emission, we can consult Figure~7 in \citet{tatischeff02},
which illustrates the X-ray production rate by the low energy electron impact. The
injected electrons are assumed to have a total energy rate of 1~erg~s$^{-1}$ in an
energy range of 10--100~keV and the power-law number density with an index of 2. At
10~keV, a continuum photon production rate of $\sim$8$\times$10~s$^{-1}$~keV$^{-1}$ is
expected. We observed $\sim$1.1$\times$10$^{-5}$~s$^{-1}$~cm$^{-2}$~keV$^{-1}$ at 10~keV
in the XIS spectrum, which can be converted to
$\sim$9.5$\times$10$^{40}$~s$^{-1}$~keV$^{-1}$ at a distance of 8.5~kpc. Therefore, an
electron injection rate of $\sim$1$\times$10$^{39}$~erg~s$^{-1}$ is required. Assuming
that the diffuse medium is ``optically-thick'' to the electron beam, we can derive the
required flux of the injected electrons to be $\sim 2 \times
10^{1}$~erg~s$^{-1}$~cm$^{-2}$ regardless of the target density. Integrating over the
entire mass of the medium \citep{yusef-zadeh02b,wang06} may lead to a significant
overestimation.

To summarize, the electron ionization interpretation requires the following conditions:
(1) the electron beam has an incident flux of $\sim 2 \times 10^{1}$
erg~s$^{-1}$~cm$^{-2}$ and (2) an index of $\sim$2 for the injected power-law number
density, and (3) 4-5 times larger abundance than solar for iron in the diffuse medium.

We do not have a list of electron accelerators with an estimated energy injection rate
in the Galactic center region and further studies must be undertaken to elucidate the
primary source. However, we can at least claim that the accelerated electrons by
wind-wind collisions in the Arches cluster \citep{wang06} are unlikely the case. This is
simply because the total kinetic energy rate via stellar winds is smaller by an order
than the required electron energy injection rate. Here the total energy rate is
calculated from an assumed wind velocity of $\sim$1000~km~$^{-1}$ and a mass loss rate
of $\sim$10$^{-4}$~$M_{\odot}$~yr$^{-1}$ estimated by the free-free emission intensity
at the centimeter continuum \citep{lang05}.

Protons accelerated via wind-wind collisions to have an MeV energy may cause both the
bremsstrahlung and the iron line emission in a similar manner as electrons with several
tens of keV. However, because the bremsstrahlung conversion rate for protons is as
inefficient as that of electrons \citep{uchiyama02}, the protons ionization will
encounter the same energy budget deficit.

Finally, heavy ions with an MeV~amu$^{-1}$ energy are also capable of producing the
continuum emission by inverse bremsstrahlung and the iron line emission by K shell
ionization. In this case, however, the resultant lines are broader and bluer than the
case of proton or electron ionization by $\sim$50~eV (e.g., \cite{burch71}), which
disagrees with the iron K$\alpha$ line center determined from the XIS
spectrum. Therefore, the ionization by heavy ions can be excluded regardless of whether
they are accelerated by wind-wind collisions in the Arches cluster or somewhere else in
the Galactic center region.

\section{Summary}\label{s6}
We conducted a spectroscopy study of the hard X-ray emission in the Arches cluster using
the Suzaku XIS data during the performance verification phase. We obtained a high
signal-to-noise spectrum in the 3.0--12~keV range, which consists of a thermal (\kt
$\sim$ 2.2~keV), a power-law ($\Gamma \sim 0.7$), and two Gaussian line ($E \sim$6.4 and
$\sim$7.1~keV) components. The thermal emission is from an ensemble of point sources
plus compact thermal diffuse emission at the center of the Arches cluster, while the
power-law and the Gaussian components are from the more extended diffuse medium
associated with the cluster. The $\sim$6.4 and $\sim$7.1~keV emission lines are
considered to be K$\alpha$ and K$\beta$ lines from neutral iron.

From the band-limited images, we found a similarity in the spatial distribution between
the 6.4~keV and 7.5--10.0~keV emission across the XIS field, including the local excess
at the Arches cluster. This strongly suggests that the power-law emission is related to
the 6.4 and 7.1~keV line emission in its origin.

Two ideas are examined to account for both the continuum and line emission; the
photoelectric ionization in which the Thomson scattering and the fluorescence are
respectively responsible for the continuum and the line emission, and the electron
impact ionization in which the bremsstrahlung and the K shell vacancy filling are
respectively for the continuum and the lines. Whichever the case, the Arches cluster is
unlikely the primary source.

We have shown that the combined measurements of both the line and the continuum emission
give a tight constraint on the primary source. There are dozens of $\sim$6.4~keV clumps
in the Galactic center region, and the present results in the Arches cluster illustrate
that Suzaku XIS spectroscopy is a powerful tool for characterizing the K$\alpha$ and
K$\beta$ lines as well as the underlying power-law emission. Another report of a
$\sim$6.4~keV emitter can be found in this volume \citep{koyama06c}. All these new
pieces of evidence on both types of emission should eventually be combined to reach a
unified picture of this curious phenomenon.

\bigskip

The authors express gratitude to the diligent work by the Suzaku team and the helpful
comments on the draft by Richard L. Kelly. M.\,T. acknowledges the hospitality at the
Department of Physics, Kyoto University during the course of this work. M.\,T. and
Y.\,H. are financially supported by the Japan Society for the Promotion of Science.

\end{document}